\documentclass[draftcls,onecolumn,12pt]{IEEEtran} 
\usepackage{epsfig}
\usepackage{color}
\usepackage{amsmath}
\usepackage{amssymb}
\usepackage{cite}

\newtheorem{theorem}{Theorem}

\newtheorem{defn}{Definition}
\newtheorem{ex}{Example}
\newtheorem{property}{Property}
\newtheorem{corollary}{Corollary}

\def\corr[#1]{\par\indent{\em Corollary (to #1): }}

\def\Pr{{\mathrm{Pr}}}
\def\ig{{\mathrm{IG}}}
\def\gig{{\mathrm{GIG}}}

\def\QED{\mbox{\rule[0pt]{1.5ex}{1.5ex}}}
\def\proof{\noindent\hspace{2em}{\it Proof: }}

\begin{document}

\title{Molecular Communication in Fluid Media: The Additive Inverse Gaussian Noise Channel}

\author{
K. V. Srinivas, Raviraj S. Adve, and Andrew W. Eckford%
\thanks{K. V. Srinivas and Raviraj S. Adve are with The Edward S. Rogers Sr.
Dept. of Electrical and Computer Engineering, University of Toronto,
10 King's College Road, Toronto, Ontario, Canada M5S 3G4.
Emails: \{kvsri, rsadve\}@comm.utoronto.ca}%
\thanks{Andrew W. Eckford is with the
Department of Computer Science and Engineering, York University,
4700 Keele Street, Toronto, Ontario, Canada M3J 1P3.
Email: aeckford@yorku.ca}%
}
\maketitle


\begin{abstract}

We consider molecular communication, with information conveyed in the
time of release of molecules. The main contribution of this paper is
the development of a theoretical foundation for such a communication
system. Specifically, we develop the additive inverse Gaussian (IG)
noise channel model: a channel in which the information is corrupted by
noise with an inverse Gaussian distribution. We show that such a
channel model is appropriate for molecular communication in fluid media
- when propagation between transmitter and receiver is governed by
Brownian motion and when there is positive drift from transmitter to
receiver. Taking advantage of the available literature on the IG
distribution, upper and lower bounds on channel capacity are developed,
and a maximum likelihood receiver is derived. Theory and simulation
results are presented which show that such a channel does not have a
single quality measure analogous to signal-to-noise ratio in the AWGN
channel. It is also shown that the use of multiple molecules leads to
reduced error rate in a manner akin to diversity order in wireless
communications. Finally, we discuss some open problems in molecular
communications that arise from the IG system model.

\end{abstract}


\IEEEpeerreviewmaketitle

\section{Introduction}

Modern communication systems are almost exclusively based on the
propagation of electromagnetic (or acoustic) waves. Of growing recent
interest, nanoscale networks, or nanonetworks, are systems of
communicating devices, where both the devices themselves and the gaps
between them are measured in nanometers~\cite{bush-book}. Due to the
limitations on the available size, energy, and processing power, it is
difficult for them to communicate through conventional means such as
electromagnetic or acoustic waves. Thus, communication between
nanoscale devices will substantially differ from the well known
wired/wireless communication scenarios.

In this paper, we address communication in a nanonetwork operating in a
aqueous environment; more precisely, we consider communication between
two nanomachines connected through a fluid medium, where messages are
encoded in patterns of molecules. In this scheme, the transmitter sends
information to the receiver by releasing molecules into the fluid
medium connecting them; the molecules propagate through the fluid
medium; and the receiver, upon receiving the molecules, decodes the
information by processing or reacting with the molecules. This method,
known as {\em molecular communication} \cite{hiy05}, is inspired by
biological micro-organisms which exchange information through
molecules. Information can be encoded on to the molecules in different
ways, such as using timing, concentration, or the identities of the
molecules themselves.

Molecular communication has recently become a rapidly growing
discipline within communications and information theory. The existing
literature that can be divided into two broad categories: in the first
category, components and designs to implement molecular communication
systems are described; for example, communications based on calcium ion
exchange~\cite{nak05} and liposomes~\cite{mor06} have been proposed.
These are commonly used by living cells to communicate. Other work
(e.g.,~\cite{moo06,hiy08}) has explored the use of molecular motors to
actively transport information-bearing molecules. To date, a
considerable amount of work has been done in related directions, much
of which is beyond the scope of this paper; a good review is found
in~\cite{hiy10}.

In the second category, channel models are analyzed and
information-theoretic capacity obtained, largely via simulations. Our
own prior work falls in this category: in~\cite{eck-arxiv}, idealized
models and mutual information bounds were presented for a Wiener
process model of Brownian motion without drift; while
in~\cite{kad09,kae-submit}, a net positive drift was added to the
Brownian motion and mutual information between transmitter and receiver
calculated using simulations. Aside from our own work, mutual
information has been calculated for simplified transmission models
(e.g., on-off keying) in~\cite{moo09,ata07}; while communication
channel models for molecular \emph{concentration} have been presented
in~\cite{pie10}, and mutual information calculated in~\cite{tho04}.
Less closely related to the current paper, information-theoretic work
has also been done to evaluate multiuser molecular communication
channels~\cite{ata09}, and evaluate the capacity of calcium relay
channels~\cite{nak10}.  Related work also includes
information-theoretic literature on the trapdoor
channel~\cite{bla61,per08}, and the queue-timing
channel~\cite{ana96,sun06}.

Building on the work in~\cite{kae-submit}, in this paper, we consider a
molecular timing channel in the presence of Brownian motion with
positive drift. Brownian motion is physically realistic for
nanodevices, since these devices have dimensions broadly on the same
scale as individual molecules; and we choose positive drift since it
arises in our applications of interest (e.g., communications that takes
advantage of the bloodstream). Our focus here is on the channel; we
assume that the transmitter and receiver work perfectly. We assume the
receiver has infinite time to guarantee that all transmitted molecules
will arrive and that there are no ``stray'' particles in the
environment. Therefore, in our system, communication is corrupted only
by the inherent randomness due to Brownian motion.

The key contributions of this paper are:
\begin{itemize}
    \item Most importantly, we show that a molecular timing channel
        can be abstracted as an additive noise channel with the
        noise having {\em{inverse Gaussian}} (IG) distribution
        (Section \ref{sec:ChannelModel}); thus, the molecular
        communication is modeled as communication over an
        \textit{additive inverse Gaussian noise} (AIGN) channel.
        This forms the basis of the theoretical developments that
        follow.

    \item Using the AIGN framework, we obtain upper and lower
        bounds on the information theoretic capacity of a molecular
        communication system (Theorem \ref{thm:capbound}).

    \item We investigate receiver design for molecular
        communication and present three key results: A maximum
        likelihood estimator (Theorem \ref{thm:MLEstimate})   and
        an upper bound on the symbol error probability
        (Theorem~\ref{thm:PEBound}).  We also show an effect
        similar to diversity order in wireless communications when
        multiple molecules are released simultaneously
        (Theorem~\ref{thm:UpperBoundAsymptotic}).
\end{itemize}
While the work in~\cite{kae-submit} is based largely on simulations,
the AIGN framework developed here allows us to place molecular
communications on a theoretical footing. However, we emphasize that
this paper remains an initial investigation into the theory of
molecular communications in fluid media.

This paper is organized as follows: Section~\ref{sec:ChannelModel}
presents the system and channel model under consideration.
Section~\ref{sec:capbounds} then uses this channel model to develop
capacity bounds for this system. Section~\ref{sec:RxDesign} then
develops a maximum likelihood (ML) receiver. Section~\ref{sec:conc}
wraps up the paper with extensive discussion, a few open problems and
some concluding remarks.

{\textbf{Notation:}} $h(X)$ denotes the differential entropy of the
random variable $X$. $X\sim \exp(\gamma)$ implies that $X$ is an
exponentially distributed random variable with mean $1/\gamma$, i.e.,
$f_X(x)=\gamma \exp^{-\gamma x},x>0$. $\mathcal{L}(X)$ denotes the
Laplace transform of the the probability density function (pdf) of the
random variable $X$.  Throughout the paper, $\log$ refers to the
natural logarithm, hence information is measured in nats.

\section{System and Channel model} \label{sec:ChannelModel}

Let $W(x)$ be a continuous-time random process which represents the
position at time $x$ of a molecule propagating via Brownian motion. Let
$0 \leq x_1 < x_2 < \ldots < x_k$ represent a sequence of time
instants, and let $R_i = W(x_i) - W(x_{i-1})$ represent the increments
of the random process for $i \in \{1,2,\ldots,k\}$. Then $W(x)$ is a
{\em Wiener process} if the increments $R_i$ are independent Gaussian
random variables with variance $\sigma^2 (x_i - x_{i-1})$. The Wiener
process has {\em drift} if $E[R_i] = v(x_i - x_{i-1})$, where $v$ is
the drift velocity. The Wiener process is an appropriate model for
physical Brownian motion if friction is negligible~\cite{gol82}.

\begin{figure}
\centering
	\includegraphics[width = 3.0in]{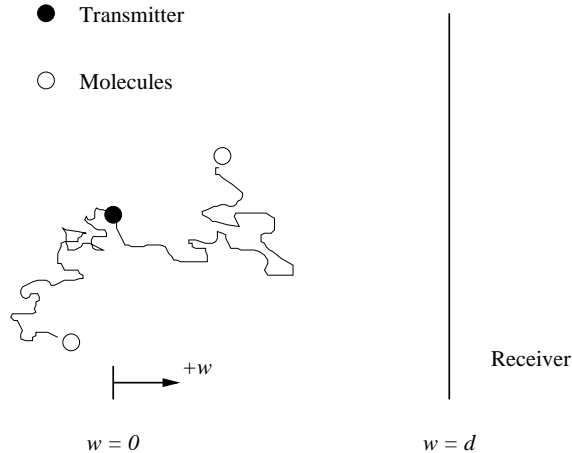}
	\caption{System Model with transmitter at $w=0$ and receiver at $w=d$}
	\label{fig:sysmodel}
\end{figure}

The system under consideration is illustrated in
Fig.~\ref{fig:sysmodel}. The transmitter releases one or more molecules
into the fluid medium at some chosen times; the molecules then
propagate to the receiver. The receiver notes the arrival time(s) and
uses this to estimate the time(s) of transmission. In the figure the
receiver is depicted as a wall, since we assume that molecules cannot
propagate beyond the receiver -- and once a molecule arrives, it is
absorbed and does not return to the medium. We therefore model
one-dimensional propagation; however, our analysis doesn't change in a
two- or three-dimensional environment, as long as the environment is
isotropic.

Consider a fluid medium with positive drift velocity $v$ and free
diffusion coefficient $D$, where the Wiener process variance is given
by $\sigma^2 = D/2$ (see footnote\footnote{In~\cite{ber-book}, values
of $D$ between 1-10 $\mu$m$^2$/s were considered realistic for
signalling molecules.}). A molecule is released into this fluid at time
$x=0$ at position $w=0$. Under the Wiener process, the probability
density of the particle's position $w$ at time $x>0$ is given
by~\cite{karatzas-book}
\begin{equation}
	f_W(w;x)
	= \frac{1}{\sqrt{2\pi \sigma^2 x}} \mathrm{exp}\left(-\frac{(w-vx)^{2}}{2\sigma^2 x}\right) .
	\label{eqn:gaussdiff}
\end{equation}
That is, treating the time $x$ as a parameter,
the pdf of the position $w$ is Gaussian with mean $vx$ and variance
$\sigma^2 x$.

Since the receiver acts as a perfectly absorbing boundary, we are only
concerned with the {\em first arrival time} $N$ at the boundary. We
assume that the transmitter is located at the origin, and in the axis
of interest, the receiver is located at position $d > 0$. In this case,
the first arrival time is given by
\begin{equation}
	N = \min \{x: W(x) = d \}.
\end{equation}
The key observation here is that if $v > 0$, the pdf of $N$, denoted by
$f_N(n)$, is given by \emph{the inverse Gaussian} (IG)
distribution~\cite{chh-book}
\begin{equation}
	\label{eqn:IG}
	f_N(n) = \left\{
		\begin{array}{cc}
			\sqrt{\frac{\lambda}{2 \pi n^3}}
			\exp \left( -\frac{\lambda (n - \mu)^2}{2 \mu^2 n} \right) , & n > 0; \\
			0, & n \leq 0 .
		\end{array}
	\right.
\end{equation}
where
\begin{eqnarray}
	\mu & = & \frac{d}{v} , \:\: \mathrm{and} \\
	\lambda & = & \frac{d^2}{\sigma^2} .
\end{eqnarray}
The mean and the variance of $N$ are given by $m_N=\mu$ and
$\mathsf{Var}(N)=\frac{\mu^3}{\lambda}$, respectively. We will use
$\ig(\mu,\lambda)$ as shorthand for this distribution, i.e., $N \sim
\ig(\mu,\lambda)$ implies (\ref{eqn:IG}).  It is important to note that
if $v = 0$, the distribution of $N$ is not IG. Furthermore, if $v < 0$,
there is a nonzero probability that the particle never arrives at the
receiving boundary. Throughout this paper, we will assume that $v
> 0$.

To develop our molecular communication channel, we assume that the
processes $W(x)$ are independent for different molecules. The
information to be transmitted is encoded in the transmit time of each
molecule. The transmitter sends symbols $X \in \mathbb{R}_{+}$, where
$\mathbb{R}_+$ represents the set of nonnegative real numbers; the
symbol $X = x$ represents a release of a single molecule at time $x$.
This molecule has initial condition $W(x) = 0$; the molecule propagates
via a Wiener process with drift velocity $v > 0$, and Wiener process
variance coefficient $\sigma^2$. This process continues until arrival
at the receiver, which occurs at time $Y \in \mathbb{R}_+$. We assume
that the propagation environment is unlimited and that, other than the
receiving boundary, nothing interferes with the free propagation of the
molecule. Under these assumptions, for a single molecule, clearly
\begin{equation}
	\label{eqn:additive}
	Y = X + N ,
\end{equation}
where $N$ is the first arrival time of the Wiener process.
Substituting into (\ref{eqn:IG}),
the probability of observing channel output $Y=y$ given
channel input $X=x$ is given by
\begin{equation}
	\label{eqn:ChannelPDF}
	f_{Y|X}(y|x) =
	\left\{
		\begin{array}{cc}
			\sqrt{\frac{\lambda}{2 \pi (y-x)^3}}
			\exp \left( -\frac{\lambda (y - x - \mu)^2}{2 \mu^2 (y-x)} \right) , & y > x; \\
			0, & y \leq x .
		\end{array}
	\right.
\end{equation}
It is apparent that the channel is affected by additive noise, in the
form of the random propagation time $N$; furthermore, by assumption,
this is the only source of uncertainty or distortion in the system. As
the additive noise $N$ has the IG distribution, we refer to the channel
defined by (\ref{eqn:additive})-(\ref{eqn:ChannelPDF}) as an
\emph{additive inverse Gaussian noise channel}. Note that we assume
that the receiver can wait for infinite time to ensure that the
molecule does arrive.

The results below follow directly from this IG framework. Several of
the results are based on properties of the IG distribution available
in~\cite{chh-book}. Previous works on the IG distribution were
motivated by its application in diverse fields such as financial,
reliability, hydrology, linguistics and demography~\cite{chh-book,
ses-book}.

\section{Capacity Bounds}
\label{sec:capbounds}

\subsection{Main Result}

Equation~\eqref{eqn:additive} is reminiscent of the popular additive
white Gaussian noise (AWGN) channel, a crucial parameter of which is
the channel capacity. As in the
AWGN case, the mutual information between the input and the output of
the channel is given by
\begin{eqnarray}
\nonumber I(X;Y) & = & \nonumber h(Y)-h(Y \vert X), \\
                 & = & \nonumber h(Y)-h(X+N\vert X) = h(Y)-h(N\vert X), \\
                 & = & h(Y)-h(N), \label{eqn:ixy1}
\end{eqnarray}
since $X$ and $N$ are independent. The capacity of the channel is the
maximum mutual information, optimized over all possible input
distributions $f_X(x)$. The set of all possible input distributions is
determined by the constraints on the input signal $X$. With the
information being encoded in the release time of the molecule, there is
no immediate analog to input power for the AWGN channel; the
constraints are application dependent, e.g., both peak-constrained and
mean-constrained inputs appear reasonable. So far, peak constraints
have not been analytically tractable; in this paper we constrain the
mean of the input signal such that
\begin{equation}
\label{eqn:inputConstr}
E[X] \leq m.
\end{equation}
That is, on average we are only willing to wait $m$ seconds to transmit our signal.
Thus, we define capacity as follows:
\defn {The capacity of the AIGN channel with input $X$ and mean constraint $E[X] \leq m$ is
defined as
\begin{equation}
\label{eqn:cap2}
C = \max_{f_X(x):E[X] \leq m} I(X;Y).
\end{equation}
}

From the receiver's perspective, $E[N]$ is finite as long as $v > 0$,
so~\eqref{eqn:inputConstr} ensures that the expected time of arrival at
the receiver is constrained, i.e., $E[Y] =E[X]+E[N] \leq m + E[N]$.
Further, note that peak constraints are not possible at the receiver,
since the pdf of $N$ is supported on $[0,\infty)$.

Unfortunately, unlike the AWGN channel, there is no simple closed-form,
single-parameter characterization of the AIGN channel capacity;
however, we use the IG distribution to form bounds on the capacity.
Thus, our main result in this section is an upper and lower bound on
the capacity of the AIGN channel.

Prior to stating this result, we need the following two properties of
the IG distribution:
\begin{property}[Differential Entropy of the IG distribution]
\label{property:DiffEntropy} Let $h_{\text{IG}(\mu,\lambda)}$ represent
the differential entropy of the IG distribution with the parameters
$\mu$ and $\lambda$.  Then
\begin{equation}
	\label{eqn:DiffEntropy}
	h_{\text{IG}(\mu,\lambda)} = \log \left(2K_{-1/2}(\lambda/\mu)\mu \right) +
          \frac{3}{2}\frac{\frac{\partial}{\partial\gamma}K_{\gamma}(\lambda/\mu)
          \left\vert_{\gamma=-1/2}\right.}{K_{-1/2}(\lambda/\mu)}+
          \frac{\lambda}{2\mu}\frac{K_{1/2}(\lambda/\mu)+
          K_{-3/2}(\lambda/\mu)}{K_{-1/2}(\lambda/\mu)} ,
\end{equation}
where $K_\gamma (\cdot)$ is the order-$\gamma$ modified Bessel function
of the third kind. \hfill $\blacksquare$
\end{property}
This property is easily derived from the differential entropy of a
generalized IG distribution; see Appendix~\ref{sec:DiffEntropy}. An
expression for the derivative of the Bessel function with respect to
its order, needed in the second term of~\eqref{eqn:DiffEntropy}, is
given in~\cite{bry05}.
\begin{property}[Additivity property of the IG distribution, from~\cite{chh-book}]
\label{property:Additivity} Let $N_i\sim
\ig(\mu_i,\lambda_i),i=1,\ldots,l,$ be $l$ not necessarily independent
IG random variables and $\frac{\lambda_i}{c_i\mu_i^2}=\kappa$ for all
$i$, and let $N=\sum_i c_i N_i$, $c_i>0$.  Then $N\sim \ig(\sum_i
c_i\mu_i,\kappa(\sum_i c_i\mu_i)^2)$. \hfill $\blacksquare$
\end{property}

The bounds on the capacity $C$ are then given by the following theorem.
\theorem {\label{thm:capbound} The capacity of the AIGN channel,
defined in~\eqref{eqn:cap2}, is bounded as
\begin{equation}
\label{eqn:thm1}
h_{\text{IG}(m+\mu,(\lambda/\mu^2)(m+\mu)^2)} - h_{\text{IG}(\mu,\lambda)} \leq C
                        \leq \log ((\mu+m)e) - h_{\text{IG}(\mu,\lambda)},
\end{equation}
where $h_{\text{IG}(\mu,\lambda)}$ is given by
Property~\ref{property:DiffEntropy}.} \proof From~\eqref{eqn:ixy1},
\begin{equation}
I(X;Y)=h(Y)-h_{\text{IG}(\mu,\lambda)},
\end{equation}
with $h_{\text{IG}(\mu,\lambda)}$ given by
Property~\ref{property:DiffEntropy}. $I(X;Y)$ is therefore maximized by
maximizing $h(Y)$ subject to the constraint given
by~\eqref{eqn:inputConstr}, equivalently $E[Y] \leq m+\mu$. Hence,
$I(X;Y)$ achieves its maximum value when $h(Y)$ is maximized subject to
the following two constraints: first, $f_Y(y) = 0, y<0$, and second,
$E[Y] \leq m+\mu$.

For the upper bound, for a random variable with a mean constraint, it
is known that the exponential distribution, defined over the interval
$(0,\infty)$, is the entropy maximizing distribution~\cite{cov-book}.
Let $\tilde{Y} \sim \exp(1/m+\mu))$; then $h(\tilde{Y})=\log((m+\mu)e)
\geq h(Y)$ for any possible distribution of $Y$ with $E[Y] = m+\mu$.
Thus,
\begin{equation}
\label{eqn:capUB}
C \leq \log ((m+\mu)e) - h_{\text{IG}(\mu,\lambda)}.
\end{equation}

 For the lower bound, suppose the input signal $X$ is IG distributed
with mean equal to $m$, satisfying~\eqref{eqn:inputConstr}. Choose the
second parameter of the IG distribution for the input signal $X$ as
$(\lambda/\mu^2) m^2$ i.e., $X \sim \ig(m, (\lambda/\mu^2) m^2)$. Then
from Property~\ref{property:Additivity}, $Y \sim
\ig(m+\mu,(\lambda/\mu^2) (m+\mu)^2)$ and
$h(Y)=h_{\text{IG}(m+\mu,(\lambda/\mu^2) (m+\mu)^2)}$. The mutual
information is given by
\begin{equation}
I(X;Y) = h_{\text{IG}(m+\mu,(\lambda/\mu^2) (m+\mu)^2)} - h_{\text{IG}(\mu,\lambda)}
\end{equation}
Note that $f_Y(y)$ in this case is not necessarily an entropy maximizing
distribution for a given mean of $m+\mu$, and hence
\begin{equation}
\label{eqn:capLB}
C \geq h_{\text{IG}(m+\mu,(\lambda/\mu^2) (m+\mu)^2)} - h_{\text{IG}(\mu,\lambda)} .
\end{equation}
The theorem follows from (\ref{eqn:capUB}) and (\ref{eqn:capLB}). \hfill \QED

Note that if one could find a valid pdf for $X$ (with $E[X] \leq m$)
that resulted in an exponential distribution for $Y$ (via convolution
with the IG distribution of $N$) then the expression
in~\eqref{eqn:capUB} would be the true capacity for mean constrained
inputs. For example, at asymptotically high velocities, i.e., as $v
\rightarrow \infty$, $\mu = d/v \rightarrow 0$ and the variance
$\mathsf{Var}(N)=\frac{\mu^3}{\lambda} \rightarrow 0$, i.e., the noise
distribution tends to the Dirac delta function. The fact that
$\frac{N}{\mu} \rightarrow 1$ as $v \rightarrow \infty$ is proven
in~\cite{ses-book}. The fact that $Y$ is distributed exponentially then
leads to the conclusion that, at high drift velocities, the optimal
input $X$ is also exponential, i.e., $X \sim \exp(1/m)$.

At low velocities, the situation is considerably more complicated. As
shown in Appendix~\ref{sec:LowVelocities}, the deconvolution of the
output ($Y$) and noise ($N$) pdfs leads to an invalid pdf, i.e., at
asymptotically low velocities, this upper bound does not appear
achievable.

\subsection{Numerical Results}
\label{sec:numResults}

We now present numerical results by evaluating the mutual information
of the AIGN channel and, in order to illustrate the upper and lower
bounds, we consider four cases:
\begin{enumerate}
\item $Y\sim \exp(1/(m+\mu))$,
\item $Y\sim \ig(m+\mu,(\lambda/\mu^2) (m+\mu)^2)$,
\item $X$ is uniformly distributed in the range $[0,2m]$,
\item $X$ is exponentially distributed with mean $m$, i.e., $X \sim \exp(1/m)$
    with $v \geq \sqrt{2\sigma^2/m}$. The need for this constraint
    is explained below.
\end{enumerate}
In all the four cases, $m=1$.
The first two choices correspond to the upper and lower bounds in
Theorem \ref{thm:capbound}, respectively. The final two choices also
provide lower bounds on the capacity, though in these cases we can only
express $f_Y(y)$ (and not $h(Y)$) in closed form; numerical integration
must be used to calculate mutual information. In the case where $X$ has
the uniform distribution on $[0,2m]$, convolving the input and noise
distributions leads to
\begin{equation}
	\label{eqn:Ypdf}
	f_Y(y) =
	\left\{
		\begin{array}{ll}
			\frac{1}{2m} F_N(y), & y \leq 2m; \\
			\frac{1}{2m} \left(F_N(y)-F_N(y-2m) \right), & y > 2m .
		\end{array}
	\right.
\end{equation}
where $F_N(n)$ is the cumulative distribution function (cdf) of $N$ and
is given by~\cite{chh-book}
\begin{equation}
\label{eqn:igcdf}
F_N(n)=\Phi\left(\sqrt{\frac{\lambda}{n}}\left(\frac{n}{\mu}-1\right) \right) +
e^{2\lambda/\mu} \Phi\left(-\sqrt{\frac{\lambda}{n}}\left(\frac{n}{\mu}+1\right) \right)
\end{equation}
where $\Phi(z)=\frac{1}{2}\left(1+
\mbox{erf}\left(\frac{z}{\sqrt{2}}\right) \right)$ is the cdf of a
standard Gaussian distributed random variable $Z$. In the case where $X
\sim \exp(1/m)$ with $m > 2\sigma^2/v^2$, the convolution leads
to~\cite{sch02}
\begin{equation}
f_Y(y)=\frac{1}{m} e^{\left(-\frac{y}{m}+d \frac{v}{\sigma^2}\right)}
    \left( e^{-kd/\sigma^2} \Phi\left( \frac{ky-d}{\sigma \sqrt{y}} \right) +
    e^{kd/\sigma^2} \Phi \left( -\frac{ky+d}{\sigma \sqrt{y}} \right) \right)
\end{equation}
where $k=\sqrt{v^2-\frac{2\sigma^2}{m}}$. The constraint on velocity,
$v^2 > 2\sigma^2/m$ ensures real $k$.

\begin{figure}
\centering
\epsfig{figure=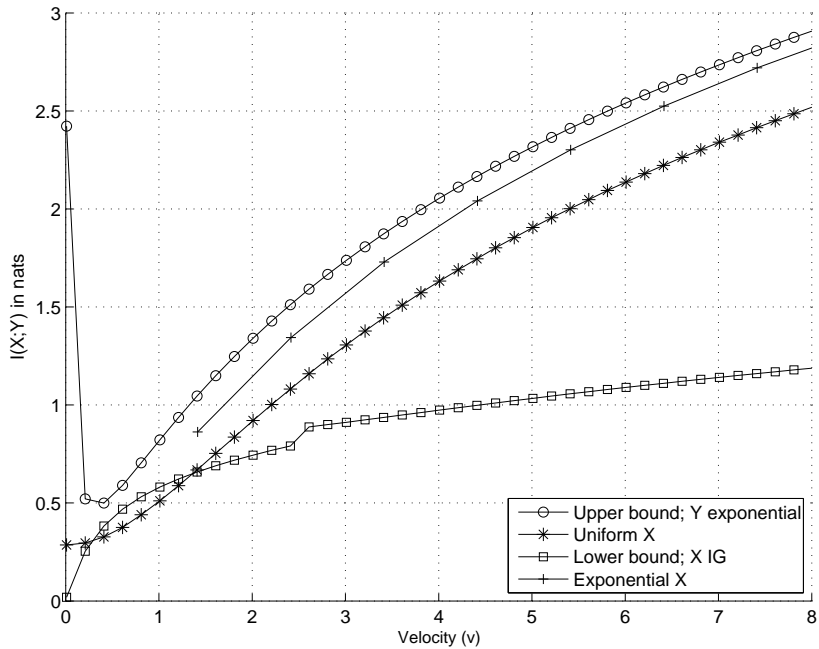,width=110mm}
\caption{Mutual information as a function of velocity; $\sigma^2 = 1$.}
\label{fig:MI4CasesVelocity}
\end{figure}

Figure~\ref{fig:MI4CasesVelocity} plots the mutual information as a
function of velocity for the four cases listed above. The upper and IG lower
bound are close to each other only over a narrow range of velocities. Further, the
cases with exponential and uniform inputs track the upper bound, with
the exponential input approaching the bound at high velocities. This is
consistent with the discussion in the previous section. However, given
its finite support, a uniform input may be closer to a practical
signalling scheme. Unsurprisingly, the plot shows that velocity is an
indicator of channel quality in that the mutual information increases
without bound as velocity increases. As a caveat, this understanding
may be valid only at higher velocities; the upper bound is not
monotonic, and at very low velocities the the upper bound actually
decreases with increasing velocity.

\begin{figure}
\centering
\epsfig{figure=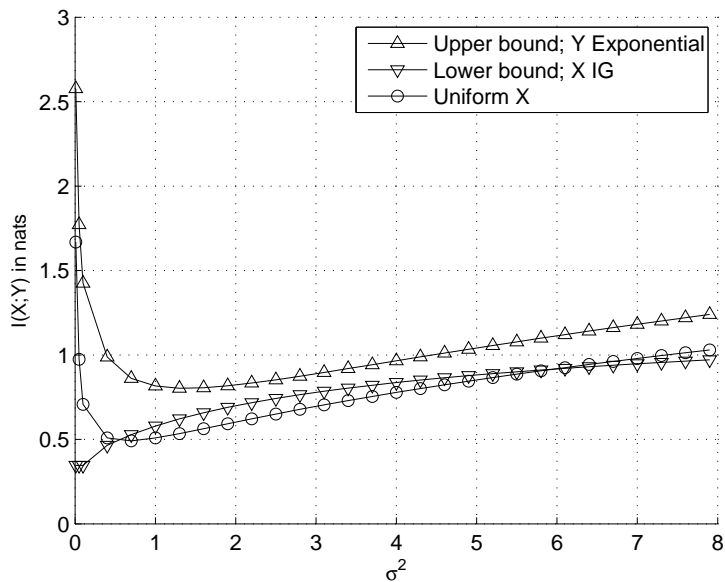,width=110mm}
\caption{Mutual information as a function of diffusion constant, $\sigma^2$ ($v=1$).}
\label{fig:UBLBUniformvsSig}
\end{figure}

The complicated relationship between mutual information and velocity
arises because, unlike AWGN channels, there is no single parameter like
SNR that determines the mutual information. The pdf in \eqref{eqn:IG}
is a function of both velocity (via $\mu$) and diffusion constant,
$\sigma^2$ (via $\lambda$). An example of this complex relationship is
shown in Fig.~\ref{fig:UBLBUniformvsSig}, where $v=1$. Both the upper
bound and the mutual information with uniform inputs fall with
increasing diffusion (randomness), but then further increasing
diffusion increases mutual information.

The increase in mutual information as a function of diffusion is
counterintuitive since diffusion is assumed to be the source of
randomness. To understand this result it is instructive to consider the
zero-velocity (no drift) case. Without diffusion, the molecule would
remain stationary at the receiver, never arriving at the receiver, and
result in zero mutual information. In this case, increasing diffusion
\emph{helps} communication. So, while it is true that diffusion
increases randomness, its impact is not monotonic.
\begin{figure}
\centering
\epsfig{figure=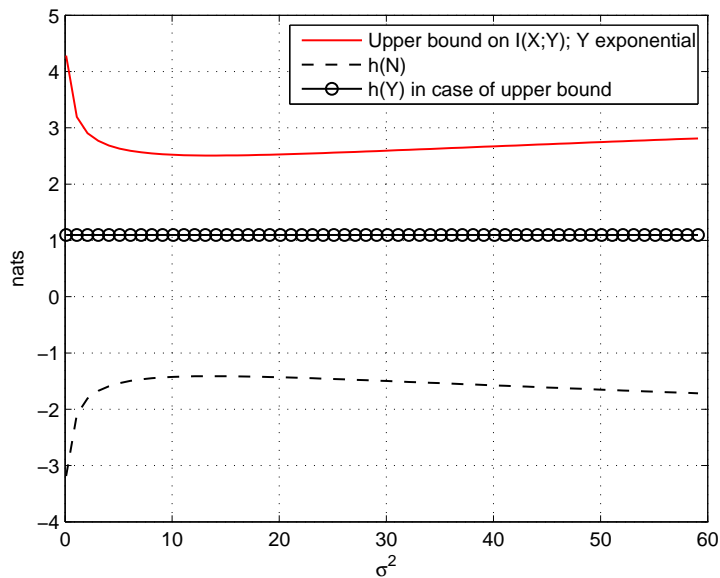,width=110mm}
\caption{Mutual information as a function of diffusion constant, $\sigma^2$ ($v=10$).}
\label{fig:UBvsSigma2V10}
\end{figure}
To illustrate this effect, consider Fig.~\ref{fig:UBvsSigma2V10}. Here,
the velocity is set relatively high ($v = 10)$. The plots are the
entropies and mutual information (upper bound) as a function of the
diffusion constant. Here, the upper bound falls steeply until $\sigma^2
\sim 4$, very slowly until $\sigma^2 \sim 10$ and then \emph{rises}
slowly for increasing $\sigma^2$. This is because for relatively large
values of $\sigma^2$, this velocity appears ``low" and increasing
diffusion increases mutual information. This is confirmed by the
falling entropy of the noise term ($h(N)$).

To summarize, in this section we developed capacity bounds for the AIGN
channel based on the IG distribution of the molecule propagation time.
While increasing velocity increases mutual information, increasing
diffusion beyond a point also increases mutual information. Unlike the
AWGN channel, no single parameter captures the performance of the AIGN
channel.

\section{Receiver Design}
\label{sec:RxDesign}

We now discuss receivers for this channel by recovering the transmitted
message (transmission time) from the times the molecules are received.
We develop both the maximum likelihood (ML) estimator and the ML
detector, and provide an error probability analysis for the ML
detection.

\subsection{Maximum Likelihood Estimator (MLE)}
\label{sec:mle}

The ML estimator of $X$, denoted by
$\hat{X}_{\text{ML}}$, is given by
\begin{equation}
\label{eqn:mle1}
\hat{X}_{\text{ML}}=\arg \max_{t} f_{Y\vert X}(y|X = t),
\end{equation}
where
\begin{equation}
\label{eqn:mlesingle1}
f_{Y\vert X}(y|X = t)=\sqrt{\frac{\lambda}{2\pi (y-t)^3}}
    \exp\left(-\frac{\lambda}{2\mu^2} \frac{((y-t)-\mu)^2}{(y-t)}\right), y \geq t,
\end{equation}
and $f_{Y\vert X}(y|X=t) = 0$ for $y < t$. The pdf given above is
commonly known as the shifted IG distribution, or the three-parameter
IG distribution, and is denoted as $\ig(t_0,\mu,\lambda)$ where $t_0$
is the location parameter~\cite{chh-book}, or the threshold
parameter~\cite{ses-book}. The mean of the shifted IG distribution is
$\mu + t$.

\theorem {\label{thm:MLEstimate} Let $\hat{X}_{\text{ML}}$ represent
the ML estimate of the transmitted symbol $X$ in an AIGN channel. Then
\begin{equation}
\label{ival}
\hat{X}_{\text{ML}} = y + \frac{\mu^2}{\lambda} \left(\frac{3}{2} -
                    \sqrt{\frac{9}{4}+\frac{\lambda^2}{\mu^2}} \right) .
\end{equation}
}

\proof Let $\Lambda(t_i) = \log f_{Y|X}(y | X = t_i)$ represent the
log-likelihood function.  Since $\log$ is monotonic,
\begin{equation}
\nonumber
\hat{X}_{\text{ML}}=\arg \max_{t_i} f_{Y|X}(y | X = t_i) = \arg \max_{t_i} \Lambda(t_i).
\end{equation}
In our case,
\begin{equation}
\label{eqn:llf}
\Lambda(t_i)=
\left\{
\begin{array}{cl}
	-\frac{3}{2} \log (y-t_i) - \frac{\lambda}{2\mu^2} \frac{((y-t_i)-\mu)^2}{(y-t_i)}, & y>t_i, \\
         - \infty, & y \leq t_i .
\end{array}
\right.
\end{equation}
By setting $\frac{\partial \Lambda(t_i)}{\partial t} = 0$, and
searching over values of $t_i < y$, we obtain the MLE given
by~\eqref{ival}. \hfill \QED

This result is consistent with the expected high velocity case ($v
\rightarrow \infty$), wherein $\hat{X}_\textrm{ML} = y$.

\subsection{ML Detection: Symbol Error Probability Analysis}
\label{sec:sepUB}

Analogous to the use of a signal constellation in AWGN channels, we now
restrict the input to the channel, i.e, the transmission time, to take
discrete values: for $T$-ary modulation we have $X\in
\{t_1,\ldots,t_T\},0\leq t_1 < t_2, \ldots < t_T$.

Using such a discrete signal set, we analyze the error probability for
binary modulation with ML detection at the receiver. Let $X\in
\{t_1,t_2\},~0\leq t_1 < t_2$, with $\Pr(X=t_1)=p_{1}$ and
$\Pr(X=t_2)=p_{2}$. The {\em log-likelihood ratio} $L(y)$ is given by
\begin{eqnarray}
	L(y) & = & \log \frac{f(y | X = t_2)}{f(y | X = t_1)} \nonumber \\
	& = & \Lambda(t_2) - \Lambda(t_1) \nonumber \\
	\label{eqn:llr}
	& = & \left\{
	\begin{array}{cl}
		-\infty, & y \leq t_2, \\
		\frac{3}{2} \log \frac{y - t_1}{y - t_2} +
			\frac{\lambda}{2\mu^2} \left( \mu^2 \left( \frac{1}{y-t_2}-\frac{1}{y-t_1}\right) +
			t_1 - t_2 \right) , & y > t_2.
	\end{array}
	\right.
\end{eqnarray}
If $L(y)$ is positive (negative), then $t_2$ has higher (lower)
likelihood than $t_1$.  If $L(y) = 0$, then there is no preference
between $t_1$ and $t_2$; we ignore this case, which occurs with
vanishing probability.  Thus, for ML detection, the decision rule is:
\begin{center}
	Pick $X = t_2$ if $L(y) > 0$, otherwise pick $X = t_1$.
\end{center}
For MAP detection, we use the same decision rule, replacing $L(y) > 0$
with $L(y) > \log(p_1/p_2)$.

The symbol error probability (SEP) is given by
\begin{equation}
\label{eqn:sep}
P_e=p_1 \text{Pr}\{t_1 \rightarrow t_2\} + p_2 \text{Pr}\{t_2 \rightarrow t_1\},
\end{equation}
where $\Pr\{t_i \rightarrow t_j\}$ is the probability of $\hat{X}_{\text{ML}}=t_j$ when $X=t_i$.
\begin{equation}
\Pr\{t_1 \rightarrow t_2\}= \int_{y_{th}}^{\infty} f_Y(y\vert X=t_1) dy 
\end{equation}
where $y_{th}$ is the decision threshold value of $y$, satisfying $L(y_{th}) = 0$. Similarly,
\begin{equation}
\Pr\{t_2 \rightarrow t_1\}= \int_{t_2}^{y_{th}} f_Y(y\vert X=t_2) dy . 
\end{equation}
We now give an upper bound on the error probability for the case when
$p_1\geq p_2$, which is simple to calculate and yet closely
approximates the exact error probability. 
\theorem{\label{thm:PEBound} Let $X\in \{t_1,t_2\},~0\leq t_1 < t_2$, with $\Pr(X=t_1)=p_{1}$,
$\Pr(X=t_2)=p_{2}$ and $p_1 \geq p_2$. The upper bound on the symbol error probability of the ML detector in an AIGN channel with input $X$ is given by
\begin{equation}
\label{eqn:sepub2ary}
P_e < p_1 (1- F_N(t_2-t_1)).
\end{equation}
}
\begin{proof}
To prove (\ref{eqn:sepub2ary}), let
\begin{eqnarray}
	\delta & = & \int_{t_2}^{\infty} f_Y(y\vert X=t_1) dy - \int_{y_{th}}^{\infty}
                                                    f_Y(y\vert X=t_1) dy \nonumber \\
	\label{eqn:Delta}
	& = & \int_{t_2}^{y_{th}}  f_Y(y\vert X=t_1) dy .
\end{eqnarray}
Then
\begin{eqnarray}
	\Pr\{t_1 \rightarrow t_2\} & = & \int_{y_{th}}^{\infty} f_Y(y\vert X=t_1) dy \nonumber \\
	\label{eqn:t1bound}
	& = & \int_{t_2}^{\infty} f_Y(y\vert X=t_1) dy - \delta .
\end{eqnarray}
Note that $\delta > 0$ since $y_{th} > t_2$.
Furthermore,
\begin{eqnarray}
	\Pr\{t_2 \rightarrow t_1\} & = & \int_{t_2}^{y_{th}} f_{Y|X}(y | X = t_2) dy \nonumber \\
	\label{eqn:MLbound}
	 & \leq &  \int_{t_2}^{y_{th}} f_{Y|X}(y | X = t_1) dy \\
	 & = & \delta ,
\end{eqnarray}
where (\ref{eqn:MLbound}) follows since, under ML detection, $f_{Y|X}(y
| X = t_1) \leq f_{Y|X}(y | X = t_2)$ when $y \leq y_{th}$.
Finally,~\eqref{eqn:sep} becomes
\begin{eqnarray}
	P_e & = & p_1 \text{Pr}\{t_1 \rightarrow t_2\} + p_2 \text{Pr}\{t_2 \rightarrow t_1\}
                                                                                    \nonumber \\
	\label{eqn:bound-1}
	& \leq & p_1\left(\int_{t_2}^{\infty} f_Y(y\vert X=t_1) dy - \delta\right) + p_2\delta \\
	& = & p_1\int_{t_2}^{\infty} f_Y(y\vert X=t_1) dy - (p_1 - p_2)\delta \nonumber \\
	\label{eqn:bound-2}
	& \leq & p_1\int_{t_2}^{\infty} f_Y(y\vert X=t_1) dy ,
\end{eqnarray}
where the last inequality follows since $p_1 \geq p_2$ (by assumption),
and so $(p_1 - p_2)\delta$ is non-negative. Finally, note that
$\int_{t_2}^{\infty} f_Y(y\vert X=t_1) dy = 1- F_N(t_2-t_1)$,
and~\eqref{eqn:sepub2ary} follows.
\end{proof}

\corollary{The bound in~\eqref{eqn:sepub2ary} is asymptotically tight
as $v \rightarrow \infty$, i.e.,
\begin{equation}
	\label{eqn:AsymptoticallyOptimal}
	\lim_{v \rightarrow \infty} \left( P_e - p_1 (1- F_N(t_2-t_1))\right) = 0 .
\end{equation}
}

\begin{proof}
The error in bound (\ref{eqn:bound-1}) is at most $p_2 \delta$, and the
error in bound (\ref{eqn:bound-2}) is equal to $(p_1 - p_2)\delta$;
thus, the total error is at most $p_2\delta$. Noting that $\mu
\rightarrow 0$ as $v \rightarrow \infty$, we show that $\delta
\rightarrow 0$ as $\mu \rightarrow 0$. For $y \geq t_2$, we have
\begin{eqnarray}
	f_{Y|X}(y | x = t_1) & = & \sqrt{\frac{\lambda}{w\pi (y-t_1)^3}}
    \exp \left( - \frac{\lambda(y-t_1-\mu)^2}{2 \mu^2 (y-t_1)} \right) \nonumber \\
	& = & \sqrt{\frac{\lambda}{w\pi (y-t_1)^3}}
        \exp \left( - \frac{\lambda(y - t_1 - 2\mu)}{2 \mu^2} \right)
                               \exp\left(- \frac{\lambda}{2 (y-t_1)}\right) \nonumber \\
	\label{eqn:BoundLimit}
	& \leq & \sqrt{\frac{\lambda}{w\pi (t_2-t_1)^3}}
        \exp \left( - \frac{\lambda(t_2 - t_1 - 2\mu)}{2 \mu^2} \right) .
\end{eqnarray}
Finally, $\delta \rightarrow 0$ follows from substituting
(\ref{eqn:BoundLimit}) into (\ref{eqn:Delta}): since $t_2 - t_1 > 0$
(by assumption), then $f_{Y|X}(y|X = t_1) \rightarrow 0$ for all $y
\geq t_2$ as $\mu \rightarrow 0$, and (\ref{eqn:AsymptoticallyOptimal})
follows.
\end{proof}

\begin{figure}
\centering
\epsfig{figure=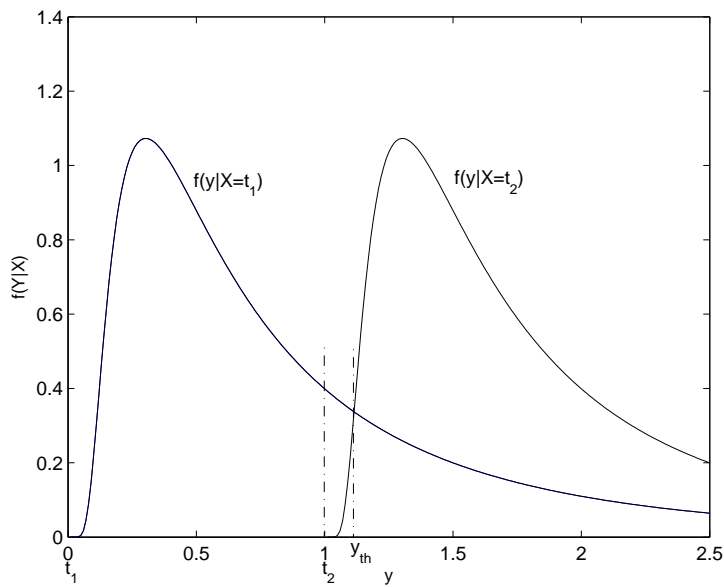,width=110mm}
\caption{Deriving the upper on symbol error probability; $t_1=0,~t_2=1,~v=1,~\sigma^2=1$
                                                                                and $d=1$.}
\label{fig:deriveUB}
\end{figure}
To illustrate this result, consider Fig.~\ref{fig:deriveUB}: $\delta$
is the area under the curve $f(y|X=t_1)$ as $y$ varies from $t_2$ to
$y_{th}$ and is always larger than $\int_{t_2}^{y_{th}} f_Y(y\vert
X=t_2) dy$, the area under the curve $f_Y(y\vert X=t_2)$ from $t_2$ to
$y_{th}$.

This bound can easily be generalized to $T$-ary modulation. When $X\in
\{t_1,\ldots,t_T\},0\leq t_1 < t_2, \ldots < t_T$ and $p_1\geq p_2 \geq
\ldots \geq p_T$, the upper bound on symbol error probability is given by
\begin{equation}
\label{eqn:sepubTary}
P_e < \sum_{i=1}^{T-1} p_i \left(1-F_N\left(t_{i+1}-t_i\right)\right).
\end{equation}
To compute the ML estimate, the receiver needs to know $\mu$ and
$\lambda$, the parameters of the noise. One way to enable the receiver
to acquire the knowledge of these parameters is by {\em{training}} as
in a conventional communication system.
Appendix~\ref{app:NoiseParameters} provides the ML estimates of these
parameters based on the IG pdf.

\subsection{Improving Reliability: Transmitting Multiple Molecules}
\label{sec:MultiMol}

The performance of a molecular communication system (the mutual
information and the error rate performance) can be improved by
transmitting multiple molecules to convey a message symbol. We assume
that the trajectories of the molecules are independent and they do not
interact with each other during their propagation from the transmitter
to the receiver.

The transmitter releases $M>1$ molecules {\em simultaneously} to convey
one of $T$ messages, $X\in\{t_1,\ldots,t_T\}$. In~\cite{kad09}, it was
shown using simulations that if multiple molecules are available,
releasing them simultaneously is the best strategy. Essentially,
releasing them at different times leads to confusion at the receiver
with molecules potentially arriving out of order. In the case of
simultaneous transmissions, the receiver observes $M$ mutually
independent arrival times
\begin{equation}
Y_j=X+N_j,~j=1,\ldots,M,
\end{equation}
where $N_j$ are i.i.d. with $N_j\sim \ig(\mu,\lambda),j=1,\ldots,M$.

\subsubsection{Maximum likelihood estimation}

We first consider ML detection of the symbol when multiple molecules
are used. Assuming that the receiver knows the values of $\mu$ and
$\lambda$ through an earlier training phase, it can use the multiple
observations $Y_j,j=1,\ldots,M,$ to obtain $\hat{X}_{\text{ML}}$.

The pdfs $f_{Y_j|X}(y_j\vert X=t_i),j=1,\ldots,M,$ are i.i.d.~with
$f_{Y_j|X}(y_j\vert X=t_i)$ given by~\eqref{eqn:mlesingle1}. The ML
estimate, in this case, is given by
\begin{align}
\hat{X}_{\text{ML}} = &\arg \max_{t_i} \prod_{j=1}^M f_{Y_j|X}(y_j\vert X=t_i) \nonumber \\
\label{eqn:mlexmulti} = & \arg \max_{t_i}
                    \prod_{j=1}^M (y_j-t_i)^{-3/2} \exp\left(-\frac{\lambda}{2\mu^2}
                    \sum_{j=1}^M \frac{((y_j-t_i)-\mu)^2}{(y_j-t_i)}\right),
                    \hspace*{0.3in} y_j>t_i.
\end{align}
Simplifying the above equation, the ML estimate can be expressed as
\begin{equation}
\label{eqn:mlemulti}
\hat{X}_{\text{ML}}=\arg \max_{t_i} \Lambda_M(t_i)
\end{equation}
where
\begin{equation}
\label{eqn:llfmulti}
\Lambda_M(t_i)=-\frac{3}{2}\sum_{j=1}^M \log (y_j-t_i) -
                        \frac{\lambda}{2\mu^2} \sum_{j=1}^M \frac{((y_j-t_i)-\mu)^2}{(y_j-t_i)},
                        \hspace*{0.3in} y_j > t_i.
\end{equation}

\subsubsection{Linear filter}

The above approach estimates the
transmitted message using a complicated ML detection filter that processes
the received signal. Given the potential applications of this research,
a simpler filter would be useful. One such filter is
the linear average, which is optimal in an AWGN channel~\cite{proakis}.
In this case, the receiver averages the $M$ observations and performs a
ML estimate with the sample mean as the test statistic. The receiver
generates
\begin{equation}
Z=\frac{1}{M}\sum_{j=1}^M Y_j.
\end{equation}
The linear filter has the following nice property: by the additivity
property of IG distribution in Property~\ref{property:Additivity},
$Z\sim \ig(E[X]+\mu,M\lambda)$. Now,
\begin{equation}
\nonumber
\hat{X}_{\text{ML}}=\arg \max_{t_i} f_Z(z\vert X=t_i),
\end{equation}
where
\begin{equation}
f_{Z|X}(z\vert X=t_i)=\sqrt{\frac{M\lambda}{2\pi (z-t_i)^3}}
        \mathrm{exp}\left(-\frac{M\lambda}{2\mu^2} \frac{((z-t_i)-\mu)^2}{(z-t_i)}\right), z>t_i.
\end{equation}
The linear receiver therefore acts as if the diffusion constant,
$\sigma^2$, is reduced by a factor of $M$ to $\sigma^2/M$. At
reasonably high velocities, this leads to better performance; however,
we have seen in Section~\ref{sec:capbounds} that, at low velocities,
diffusion can actually help communications.

At high drift velocities the reduction in the effective diffusion {\em
results in an effect akin to the diversity order in wireless
communication systems}. This is shown in the following result.
\theorem{ \label{thm:UpperBoundAsymptotic} As drift velocity $v
\rightarrow \infty$,
\begin{equation}
\label{eqn:PeBoundOneMolecule} \log(P_e) < - C_1\frac{cv^2}{\sigma^2} +
C_2 + C_3 \log{\frac{cv^2}{\sigma^2}},
\end{equation}
where $C_1, C_2$ and $C_3$ are constants. }

\begin{proof}
	The proof is found in Appendix \ref{app:UpperBoundAsymptotic}.
\end{proof}

Furthermore, for $M$ molecules and detection using the linear filter,
\begin{equation}
\label{eqn:PeBoundMMolecules} \log(P_e) < - C_1\frac{Mcv^2}{\sigma^2} +
C_2 + C_3 \log{\frac{Mcv^2}{\sigma^2}},
\end{equation}
which is essentially~\eqref{eqn:PeBoundOneMolecule} with $\sigma^2$
replaced by $\sigma^2/M$.

Since, in both~\eqref{eqn:PeBoundOneMolecule}
and~\eqref{eqn:PeBoundMMolecules}, the first term dominates at high
velocities, a semi-log plot of $P_e$ versus velocity is asymptotically
linear, with slope proportional to $- M$.

\subsection{Simulation Results}
\label{sec:rxsim}
\begin{figure}
\centering
\epsfig{figure=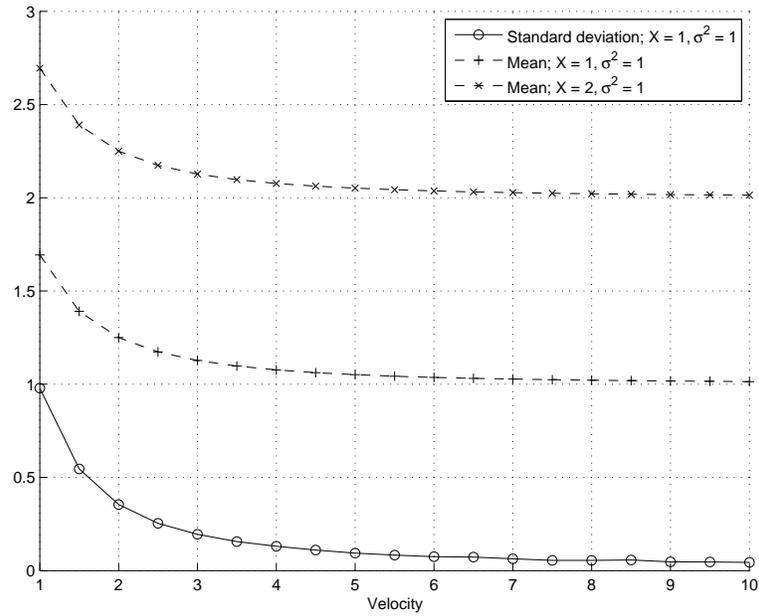,width=100mm}
\caption{Mean and standard deviation of $\hat{X}_{\text{ML}}$.}
\label{fig:mleVar}
\end{figure}

Figure~\ref{fig:mleVar} shows how the variance and the mean of the ML
estimate vary with velocity for a given $\sigma^2$. With increasing
velocity, the estimator becomes unbiased and the variance approaches
zero. As in Section~\ref{sec:capbounds}, velocity appears to be close
to the AIGN equivalent of SNR in AWGN channels; however, again, this is
only true at high velocities. At low velocities, both the velocity and
the diffusion constant play a role.

\begin{figure}
\centering
\epsfig{figure=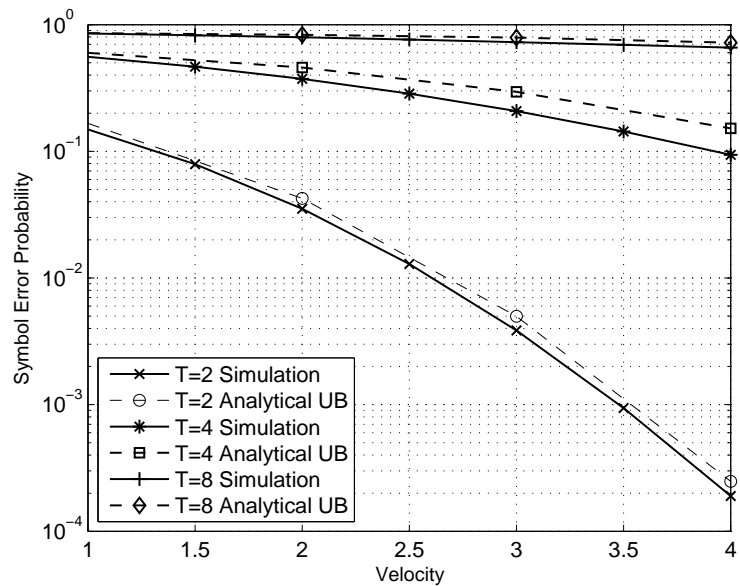,width=110mm}
\caption{Comparing the analytical upper bound and simulated error probability; single molecule
    case with $T$-ary modulation. Equiprobable symbols and $\sigma^2=1$.}
\label{fig:marySimAnal}
\end{figure}

Figure~\ref{fig:marySimAnal} plots the symbol error probability with
$T$-ary modulation for different values $T$. The input alphabet
employed for simulations is $X\in\{1+\frac{i-1}{T-1},i=1,\ldots,T\}$.
The figure also compares the upper bound on error probability,
presented in Section \ref{sec:sepUB}, with the error probability
obtained through Monte Carlo simulations. The rapidly deteriorating
error probability is clear, as is the tightness of the upper bound.

\begin{figure}
\centering
\epsfig{figure=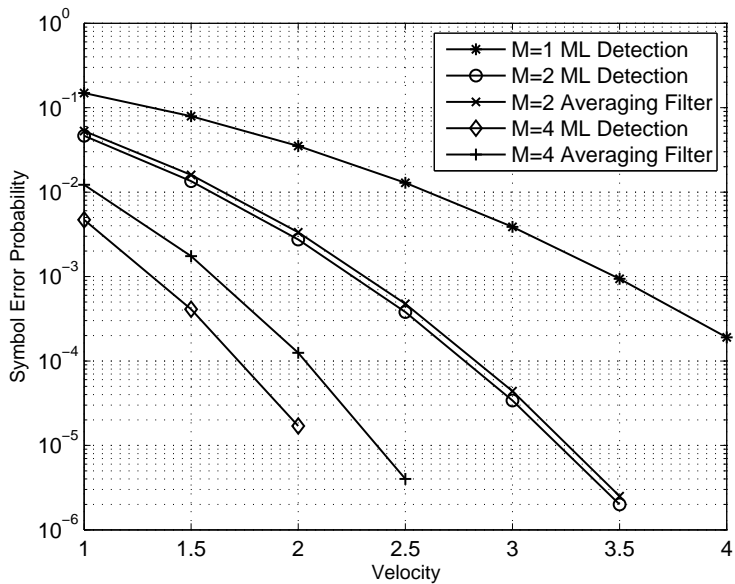,width=110mm}
\caption{Comparing the error probability of MLE with the averaging filter.
Equal a priori probabilities and $\sigma^2=1$.}
\label{fig:MultiMolAvg}
\end{figure}

The poor performance of $T$-ary modulation as shown in
Fig.~\ref{fig:marySimAnal} motivates the multiple molecule system
described in Section~\ref{sec:MultiMol}. Figure~\ref{fig:MultiMolAvg}
plots the error rate performance when $X\in \{1,2\}$ and each symbol is
conveyed by releasing multiple molecules. As expected, there is a
effect akin to receive diversity in a wireless communication system.
Here, the performance gain in the error probability increases with the
number of molecules transmitted per message symbol.

Figure~\ref{fig:MultiMolAvg} also compares the performance of the
averaging filter with the ML estimation given by~\eqref{eqn:mlemulti}.
The linear averaging filter is clearly suboptimal with performance
worsening with increasing number of molecules transmitted per symbol
($M$). This result again underlines the significant differences between
the AIGN and AWGN channel models.

\section{Discussion and Conclusions}
\label{sec:conc}

In proposing a new channel model based on IG noise, we have necessarily
analyzed the simplest possible interesting cases. In this regard, there
are several issues left unresolved.


\emph{Single versus Multiple Channel Uses}: Throughout this paper, we
have focused on the case of a {\em single channel use}, in which we use
the channel to transmit a single symbol of information; our capacity
results are measured in units of nats per channel use. Translating
these results to nats per molecule is straightforward: each channel use
consists of a deterministic number of molecules $M$, where $M \geq 1$,
thus, we merely divide by $M$. However, measuring nats per unit time is
a more complicated issue, since the duration of the channel use is a
random variable, dependent on both the input and the output.
Following~\cite{ana96}, where the capacity per unit time of a queue
timing channel was calculated with respect to the {\em average service
time}, here we can normalize our capacity results either with the {\em
average propagation time} $E[N]$, or the {\em average length of the
communication session} $E[Y]$.  Since $E[Y] = E[X]+E[N]$, our decision
to constrain the mean of the input distribution $f_X(x)$ would then
have a natural interpretation in terms of the capacity per unit time.

Further, our system model excludes the possibility of other molecules
propagating in the environment, except those transmitted as a result of
the channel use; equivalently, we assume each channel use is
orthogonal. This raises the question of how to use the channel
repeatedly: if the signalling molecules are indistinguishable, then
(under our formulation) the transmitter must wait until all $M$
molecules have arrived before a new channel use can begin.  On the
other hand, if the signalling molecules are distinguishable, then
channel uses can take place at any time, or even the same time. This is
because, if there is no ambiguity in matching received molecules to
channel uses, those channel uses are orthogonal.

\emph{Inter-symbol Interference}: Repeated channel uses also leads to a
situation akin to inter-symbol interference (ISI) in conventional
communications. Since propagation time is not bounded, the transmitter
may release the molecule corresponding to the ``next" symbol while the
``previous" molecule is still in transit. Molecules may, therefore,
arrive out of order. This problem is exacerbated if multiple molecules
are released simultaneously to achieve diversity. Decoding with such
ISI is complex since schemes such as the Viterbi algorithm cannot be
used (even ignoring the fact that the system would, in theory, have
infinite memory). This is because, in each time slot, the number of
molecules not yet received - due to transmission from previous time
slots - acts as the state of the channel with corresponding noise
distributions. In other contexts, an example of a channel with states
is the Gilbert-Elliott channel~\cite{AndrewGilbertElliott}.

\emph{Synchronization and Differential Encoding}: The system model and
the analysis presented here assumes perfect synchronization between the
transmitter and the receiver. It is unclear how difficult, or easy, it
would be to achieve this with nano-scale devices. An information
theoretic analysis of the effect of asynchronism in AWGN channels has
been presented in~\cite{tch09}. Given the importance of timing in our
model, extensions of such work to the AIGN channel would be useful. An
interesting alternative would be to use differential modulation schemes
such as \emph{interval modulation} presented in~\cite{muk02}.

\emph{Amplitude and Timing Modulation}: The work presented here focuses
on timing modulation, which leads naturally to the AIGN channel model.
A more sophisticated scheme would be to use ``amplitude" modulation as
well - such as by varying the number of molecules released. It may be
possible to leverage work on positive-only channels such as in
optics~\cite{SteveHranilovicChannelCapacity}. Amplitude modulation
could be coupled with the timing modulation considered here. However,
it is important to note that any amplitude information would reproduced
at the receiver faithfully since, in the model we have considered so
far, the receiver is allowed to wait for all molecules to arrive before
decoding. Therefore, to be useful, a reasonable model of amplitude
modulation must also include receiver imperfections and account for the
issue of ISI as described above.

\emph{Two-way Communication and Negative Drifts}: The AIGN channel
model is valid only in the case of a positive drift velocity. In this
regard, it does not support two-way communication between nano-devices.
With zero drift velocity, the mean transition time is unbounded, but
the probability that the molecule arrives approaches 1; with negative
drift velocities, even this arrival is not guaranteed~\cite{chh-book}.
Molecular communications with negative drift velocities remains a
completely open problem and one that is outside the scope of this
paper. In this case, the noise term is $\ig(-\mu,\lambda)$ and the IG
framework provided here may be used to analyze such a problem.

In conclusion, our results both illustrate the feasibility of molecular
communication and show that it can be given a mathematical framework.
However, our results lead to many interesting open questions, some of
which are described above. We believe our key contribution here has
been to provide this mathematical framework, making it possible to
tackle some of these problems.

\appendices

\section{Differential entropy of the IG distribution}
\label{sec:DiffEntropy}

Here we prove Property~\ref{property:DiffEntropy}. For a given $\mu$
and $\lambda$, the differential entropy of the noise $h(N)$ is fixed
and can be computed from the generalized IG distribution (GIG). The GIG
distribution is characterized by three parameters and the pdf of a
random variable $X$ distributed as GIG is given by~\cite{chh-book}
\begin{multline}
\label{eqn:gig}
f_X(x;\gamma,\mu,\lambda)=\frac{1}{2\mu^{\gamma}K_{\gamma}
                \left(\frac{\lambda}{\mu}\right)}~x^{\gamma-1}\exp
                    \left(-\frac{\lambda x^{-1}+(\lambda/\mu^2)x}{2}\right),\\
                    -\infty < \gamma < \infty, \mu > 0, \lambda \geq 0, x> 0,
\end{multline}
where $K_{\gamma}(\cdot)$ is the modified Bessel function of the third
kind of order $\gamma$. It is commonly denoted as
$\gig(\gamma,\mu,\lambda)$ and $\ig(\mu,\lambda)$ is a special case,
obtained by substituting $\gamma=-1/2$~\cite{chh-book}.

When $X \sim \gig(\gamma,\mu, \lambda)$, its differential entropy, in
nats, is given by~\cite{kaw03}
\begin{equation}
\label{eqn:hgig}
h(X)=\log \left(2K_{\gamma}(\lambda/\mu)\mu \right) - (\gamma-1)
        \frac{\frac{\partial}{\partial\gamma}K_{\gamma}(\lambda/\mu)}{K_{\gamma}(\lambda/\mu)}+
        \frac{\lambda}{2\mu}\frac{K_{\gamma+1}(\lambda/\mu)+
        K_{\gamma-1}(\lambda/\mu)}{K_{\gamma}(\lambda/\mu)}.
\end{equation}
Setting $\gamma = -1/2$, the differential entropy of $N\sim
\ig(\mu,\lambda)$ is given by
\begin{equation}
\label{eqn:hig}
h(N)= h_{\text{IG}(\mu,\lambda)} = \log \left(2K_{-1/2}(\lambda/\mu)\mu \right) +
          \frac{3}{2}\frac{\frac{\partial}{\partial\gamma}K_{\gamma}(\lambda/\mu)
          \left\vert_{\gamma=-1/2}\right.}{K_{-1/2}(\lambda/\mu)}+
          \frac{\lambda}{2\mu}\frac{K_{1/2}(\lambda/\mu)+
          K_{-3/2}(\lambda/\mu)}{K_{-1/2}(\lambda/\mu)} ,
\end{equation}
and the property follows.

\section{Evaluating optimal input distribution at low velocities}
\label{sec:LowVelocities}

If a pdf exists that leads to an exponentially distributed measured
signal $Y$, it would be the capacity achieving input distribution.
Furthermore, the pdf of the measured signal is the convolution of the
pdf of the input and that of IG noise pdf. We therefore attempt to
evaluate the optimal distribution at asymptotically low velocities by
deconvolving the known optimal distribution (exponential) of the output
$Y$ and the IG noise. The Laplace transform of the IG distribution is
given by
\begin{equation}
\label{eqn:LTofIG}
\mathcal{L}(N) = E[e^{-sX}]=\exp \left[\frac{\lambda}{\mu}
                \left(1-\sqrt{1+\frac{2\mu^2}{\lambda}s}\right)\right].
\end{equation}
For given values of $\sigma^2$ and $d$, as $v\rightarrow 0$,
$\mu\rightarrow \infty$ and $\gamma$ is fixed. In such a case,
$\mathcal{L}(N)$ can be approximated as
\begin{equation}
\label{eqn:LTofIGapprox}
\mathcal{L}(N)\approx \exp \left(-\sqrt{2\lambda s}\right).
\end{equation}

As $Y=X+N$, $\mathcal{L}(X)=\mathcal{L}(Y)/\mathcal{L}(N)$. To achieve
the upper bound on capacity, $f_Y(y)=\frac{1}{m_Y} e^{\frac{-y}{m_Y}}$,
where $m_Y=E[Y]=E[X]+\mu$ and hence
\begin{equation}
\mathcal{L}(Y)=\frac{1/m_Y}{s+(1/m_Y)} \Rightarrow
\mathcal{L}(X)=\frac{1/m_Y}{(1/m_Y)+s} \exp\left(\sqrt{2\lambda s} \right)
\label{eqn:LTofx}
\end{equation}
and the pdf of $X$ can be obtained by computing the inverse Laplace
transform $\mathcal{L}^{-1}(X)$. The inverse Laplace transform can be
computed by making use of the following Laplace transform
pair~\cite{het75}:
\begin{multline}
\label{Lappair} \mathcal{L}^{-1}\left\{ \frac{\exp(-c\sqrt{s+b})}{s-a}
\right\} = \frac{e^{at}}{2}\left( \exp\left(-c\sqrt{a+b}\right)
\text{erfc}\left(\frac{c}{2\sqrt{t}}-\sqrt{(a+b)t}\right) \right. \\
\left. + \exp\left(c\sqrt{a+b}\right)
\text{erfc}\left(\frac{c}{2\sqrt{t}}+\sqrt{(a+b)t}\right) \right),
\end{multline}
where $a,b$ and $c$ are constants. Using\eqref{Lappair}, we
obtain
\begin{multline}
\mathcal{L}^{-1}\left\{
\frac{1/m_Y}{s+(1/m_Y)}\exp(\sqrt{2\lambda}\sqrt{s}) \right\} =
\frac{(1/m_Y)e^{\frac{-1}{m_Yt}}}{2}\left( \exp\left(\jmath
\sqrt{2\lambda/m_Y}\right) \text{erfc}\left(-\sqrt{\lambda/2t}-\jmath
\sqrt{t/m_Y}\right) \right. \\ \left. + \exp\left(-\jmath
\sqrt{2\lambda/m_Y}\right) \text{erfc}\left(-\sqrt{\lambda/2t}+\jmath
\sqrt{t/m_Y}\right) \right)
\end{multline}
where
\begin{equation}
\nonumber
\text{erfc}(z)=\frac{2}{\sqrt{\pi}}\int_z^\infty e^{-z^2}dz
\end{equation}
Note that $\text{erfc}(z)$ can be evaluated for complex values of its
argument $z$ and $\text{erfc}(z^*)=(\text{erfc}(z))^*$, where $z^*$ is
the complex conjugate of $z$. Hence
\begin{equation}
\label{eqn:invLTofx}
f_X(x) = \frac{e^{\frac{-1}{m_Yx}}}{m_Y} \Re\left\{ \exp\left(\jmath \sqrt{2\lambda/m_Y}\right)
                \text{erfc}\left(-\sqrt{\lambda/2x}-\jmath \sqrt{x/m_Y}\right)\right\}.
\end{equation}
This, unfortunately, does not appear to be a valid pdf. The capacity of
the AIGN channel at low velocities is therefore, yet, unknown.

\subsection{When there is no drift}
To confirm the result in~\eqref{eqn:invLTofx}, we test the case of zero
velocity. Note that in this case, the noise is not IG; however, the
zero velocity case converges in limit to the case without drift.
Without drift, the arrival time has a pdf given by~\cite{chh-book},
\begin{equation}
\label{nodrift1}
f(t)=\sqrt{\frac{\lambda}{2\pi t^3}} \mathrm{exp}^{\frac{-\lambda}{2t}}, \hspace*{0.5in} t>0
\end{equation}
Note that $t \sim \mbox{Inverse Gamma}(1/2,\lambda/2)$. The inverse
Gamma distribution, with shape parameter $\alpha$ and scale parameter
$\beta$, is given by
\begin{equation}
\label{gd}
f(t;\alpha,\beta)=\frac{\beta^{\alpha}}{\Gamma(\alpha)}(1/t)^{\alpha+1}\mathrm{exp}(\beta/t),~t>0.
\end{equation}

Hence, the Laplace transform of the inverse Gamma distribution is
\begin{equation}
\mathcal{L}(N)=\mathcal{L}[Inv Gamma(1/2,\lambda/2)]=\frac{2(s \lambda/2)^{1/4}}
                                                {\sqrt{\pi}} K_{1/2}(\sqrt{2\lambda s}).
\end{equation}
Substituting
\begin{equation}
K_{1/2}(z)=\sqrt{\frac{\pi}{2z}}e^{-z},
\end{equation}
we get
\begin{equation}
\label{eqn:noiseLT2}
\mathcal{L}(N)=e^{-\sqrt{2\lambda s}}
\end{equation}
This results in
\begin{equation}
\label{eqn:xLT2}
\mathcal{L}(X)=\frac{1/m_Y}{s+(1/m_Y)}e^{\sqrt{2 \lambda s}}
\end{equation}

Note that~\eqref{eqn:noiseLT2} is same as~\eqref{eqn:LTofIGapprox} and
\eqref{eqn:xLT2} is same as~\eqref{eqn:LTofx}. Hence, we get
\eqref{eqn:invLTofx} when we try to obtain $f_X(x)$ by evaluating
$\mathcal{L}^{-1}(X)$.

\section{Estimating Noise Parameters}
\label{app:NoiseParameters}

To estimate the noise parameters, the transmitter releases $k$
``training" molecules at known time $t_0$. Let the receiver observe
$Y_j=t_0+N_j,j=1,2,\ldots,k,$ where $N_j\sim \ig(\mu,\lambda)$ are
i.i.d.~and the receiver knows $t_0$ a priori. The pdf's of
$(Y_j-t_0),j=1,\ldots,k$, are i.i.d. and IG distributed as given by
\begin{equation}
f_{Y_j-t_0}(y_j-t_0)=\sqrt{\frac{\lambda}{2\pi (y_j-t_0)^3}}
    \exp\left(-\frac{\lambda}{2\mu^2} \frac{((y_j-t_0)-\mu)^2}{(y_j-t_0)}\right), y_j>t_0.
\end{equation}

In general, $\infty < t_0 < -\infty$; however, in our case, $0 < t_0 <
\infty$. When $Y\sim \ig(t_0,\mu,\lambda)$, $m_Y=E[Y]=\mu+t_0$. When the
receiver knows the value of $t_0$, the ML estimates of the remaining
two parameters $\mu$ and $\lambda$ can be obtained as
\begin{equation}
\label{mlemu}
\hat{\mu}(t_0)=\bar{Y}-t_0,
\end{equation}
where $\bar{Y}=\frac{1}{k}\sum_{j=1}^k Y_j$ is the sample mean and
\begin{equation}
\label{mlelambda}
\hat{\lambda}(t_0)=\left[\frac{1}{k} \sum_{j=1}^k \left( \frac{1}{Y_j-t_0}-
                                        \frac{1}{\bar{Y}-t_0} \right) \right]^{-1}.
\end{equation}
Assuming $\mu$ and $\lambda$ does not change significantly from the
time the receiver estimates the parameters and the time of actual
communication, the receiver can obtain the ML estimate of the release
times of the molecules.

\section{Upper Bound on Asymptotic Error Rate}
\label{app:UpperBoundAsymptotic}

Here we prove Theorem \ref{thm:UpperBoundAsymptotic}. Recall that, for $2$-ary modulation with
$X\in\{t_1,t_2\},0\geq t_1\geq t_2$, the upper bound on SEP is given by
\begin{equation}
P_e < p_1 (1- F_N(t_2-t_1)).
\end{equation}
where
\begin{equation}
F_N(n)=\Phi\left(\sqrt{\frac{\lambda}{n}}\left(\frac{n}{\mu}-1\right) \right) +
e^{2\lambda/\mu} \Phi\left(-\sqrt{\frac{\lambda}{n}}\left(\frac{n}{\mu}+1\right) \right)
\end{equation}
where $\Phi(z) =\frac{1}{2}\left( 1+ \text{erf}\left(\frac{z}{\sqrt{2}}\right)\right)$ is the
cdf of a standard Gaussian distributed random variable $Z$. Here,
\begin{equation}
\text{erf}(z)=\frac{2}{\sqrt{\pi}}\int_0^z e^{-u^2}du
\end{equation}
For $z\gg 1$, $\text{erf}(z)$ can be approximated as
\begin{equation}
\label{eqn:erfhighz}
\text{erf}(z)\approx 1-\frac{e^{-z^2}}{\sqrt{\pi}z}
\end{equation}

Now, we compute $F_N(c)$, $c=t_2-t_1$, and examine its behavior as $v\rightarrow \infty$.
Recall that $\mu=\frac{d}{v}$ and $\lambda=\frac{d^2}{\sigma^2}$.
\begin{equation}
F_N(c)=\Phi\left( \sqrt{\frac{cv^2}{\sigma^2}}-\sqrt{\frac{d^2}{c\sigma^2}}\right)+
e^{2vd/\sigma^2}\Phi\left( -\sqrt{\frac{cv^2}{\sigma^2}}-\sqrt{\frac{d^2}{c\sigma^2}}\right)
\end{equation}

Consider the first term in $F_N(c)$.
\begin{equation}
\Phi\left( \sqrt{\frac{cv^2}{\sigma^2}}-\sqrt{\frac{d^2}{c\sigma^2}}\right) =
\frac{1}{2} \left(1+ \text{erf}\left(\frac{ \sqrt{\frac{cv^2}{\sigma^2}}-
\sqrt{\frac{d^2}{c\sigma^2}}}{\sqrt{2}}\right)\right)
\end{equation}
When $v\rightarrow \infty$, $\sqrt{\frac{cv^2}{\sigma^2}} \rightarrow
\infty$ and thus $\left(
\sqrt{\frac{cv^2}{\sigma^2}}-\sqrt{\frac{d^2}{c\sigma^2}}\right) \gg
1$. Hence, we use the approximation given by~\eqref{eqn:erfhighz} to
obtain
\begin{equation}
\Phi\left( \sqrt{\frac{cv^2}{\sigma^2}}-\sqrt{\frac{d^2}{c\sigma^2}}\right) \approx 1 -
\frac{1}{\sqrt{2\pi}}\frac{1}{\sqrt{\frac{cv^2}{\sigma^2}}-\sqrt{\frac{d^2}{c\sigma^2}}}
e^{\left(-\frac{cv^2}{\sigma^2}+\frac{2vd}{\sigma^2}-\frac{d^2}{c\sigma^2}\right)}
\end{equation}
Now, consider the second term in $F_N(c)$.
\begin{equation}
\Phi\left( -\sqrt{\frac{cv^2}{\sigma^2}}-\sqrt{\frac{d^2}{c\sigma^2}}\right)=\frac{1}{2}
\left( 1- \text{erf}\left(\frac{ \sqrt{\frac{cv^2}{\sigma^2}}+\sqrt{\frac{d^2}{c\sigma^2}}}
{\sqrt{2}}\right)\right)
\end{equation}
When $v\rightarrow \infty$ $\left(
\sqrt{\frac{cv^2}{\sigma^2}}+\sqrt{\frac{d^2}{c\sigma^2}}\right) \gg 1$
and, using the approximation given by~\eqref{eqn:erfhighz}, we obtain
\begin{equation}
e^{2vd/\sigma^2}\Phi\left( -\sqrt{\frac{cv^2}{\sigma^2}}-\sqrt{\frac{d^2}{c\sigma^2}}\right) \approx
\frac{1}{\sqrt{2\pi}}\frac{1}{\sqrt{\frac{cv^2}{\sigma^2}}+\sqrt{\frac{d^2}{c\sigma^2}}}
e^{\left(-\frac{cv^2}{\sigma^2}-\frac{d^2}{c\sigma^2}\right)}
\end{equation}

Hence,
\begin{equation}
F_N(c) \approx 1 - \frac{1}{\sqrt{2\pi}}\frac{1}{\sqrt{\frac{cv^2}{\sigma^2}}-\sqrt{\frac{d^2}
{c\sigma^2}}} e^{\left(-\frac{cv^2}{\sigma^2}+\frac{2vd}{\sigma^2}-\frac{d^2}{c\sigma^2}\right)}
+\frac{1}{\sqrt{2\pi}}\frac{1}{\sqrt{\frac{cv^2}{\sigma^2}}+\sqrt{\frac{d^2}{c\sigma^2}}}
e^{\left(-\frac{cv^2}{\sigma^2}-\frac{d^2}{c\sigma^2}\right)}
\end{equation}

As $e^{\left(-\frac{cv^2}{\sigma^2}-\frac{d^2}{c\sigma^2}\right)}$
decays faster than
$e^{\left(-\frac{cv^2}{\sigma^2}+\frac{2vd}{\sigma^2}-\frac{d^2}{c\sigma^2}\right)}$,
the second term in the above equation dominates the rate at which
$F_N(c)$ goes to $1$ as $v\rightarrow \infty$. At high velocities,
$F_N(c)$ can be approximated as
\begin{equation}
F_N(c)\approx 1-\frac{1}{\sqrt{2\pi}}\frac{e^{-\frac{cv^2}{\sigma^2}}}{\sqrt{\frac{cv^2}
{\sigma^2}}}
\end{equation}
Thus, at high velocities, the upper bound on SEP is given by
$\frac{1}{\sqrt{2\pi}}
\frac{e^{-\frac{cv^2}{\sigma^2}}}{\sqrt{\frac{cv^2}{\sigma^2}}}$.  The
theorem follows by taking the logarithm of this expression.

%
\renewcommand{\baselinestretch}{1}
\bibliographystyle{IEEEtran}


\renewcommand{\baselinestretch}{1.5}


\end{document}